\documentclass[a4paper,12pt]{article}

\usepackage[T1]{fontenc}
\usepackage[utf8]{inputenc}

\usepackage{amsmath,amssymb,amsfonts}
\usepackage{mathtools}

\DeclareMathOperator{\col}{col}
\DeclareMathOperator{\sat}{sat}

\usepackage{geometry}
\usepackage{titling}
\usepackage{cite}

\begin{document}

\begin{center}
    \Large \textbf{ROBUST QUADCOPTER MOTION CONTROL USING OUTPUT FEEDBACK}
\end{center}

\begin{center}
    \textbf{S.A. Kim}\\
    \textit{ITMO University}\\
    Russia, 197101, St. Petersburg, Kronverksky Ave., 49, liter A\\
    E-mail: skim@itmo.ru
\end{center}

\begin{center}
    \textbf{A.A. Pyrkin}\\
    \textit{ITMO University}\\
    Russia, 197101, St. Petersburg, Kronverksky Ave., 49, liter A\\
    E-mail: pyrkin@itmo.ru
\end{center}

\begin{center}
    \textbf{O.I. Borisov}\\
    \textit{ITMO University}\\
    Russia, 197101, St. Petersburg, Kronverksky Ave., 49, liter A\\
    E-mail: borisov@itmo.ru
\end{center}

\vspace{0.5cm}


\vspace{0.5cm}

\noindent \textbf{Abstract:} The study addresses the problem of quadcopter motion control using output feedback. By applying a geometric approach, the quadcopter model is transformed into a normal form with a time-varying gain coefficient, which is subsequently made stationary through double integration of the control input. A robust output feedback control law is synthesised based on the extended observer method.

\section{Introduction}

The motion of a quadcopter is described by a multidimensional nonlinear model that often contains parametric, structural, functional, and signal uncertainties. This significantly complicates the synthesis of control laws, making it a challenging task and attracting ongoing research interest. A common approach to quadcopter motion control is model linearization, which simplifies the control design to a certain extent. However, in practice, this often degrades transient performance due to model inaccuracies. Some solutions have been proposed that consider the nonlinear motion model, for instance, in references [1, 2]. Yet, these works use an incomplete model because the yaw angle dynamics are excluded, and the methodology for tuning the parameters of the proposed algorithm is cumbersome. The shortcomings were addressed in reference [3], where a complete quadcopter motion model for the dynamic positioning task was obtained. Based on this model, solutions were proposed that generalized the previous result [1] to the case of a varying yaw angle and simplified the control tuning procedure. Nevertheless, the algorithms derived in that work require full state measurement. Our work extends the previously obtained result [3] to the case where the derivatives of the output variables are not measurable. This is achieved by synthesizing a robust output feedback control law based on the extended observer method.

\section{Problem Statement}

Consider the quadcopter motion model

\begin{align*}
\begin{bmatrix} \dot{x} \\ \dot{y} \\ \dot{z} \end{bmatrix} &= 
\begin{bmatrix} v_x \\ v_y \\ v_z \end{bmatrix}, \\
\begin{bmatrix} \dot{v}_x \\ \dot{v}_y \\ \dot{v}_z \end{bmatrix} &=
- \begin{bmatrix} a_x & 0 & 0 \\ 0 & a_y & 0 \\ 0 & 0 & a_z \end{bmatrix}
\begin{bmatrix} v_x \\ v_y \\ v_z \end{bmatrix} +
\frac{1}{m}
\begin{bmatrix} c_\phi s_\theta c_\psi + s_\phi s_\psi & 0 \\ s_\phi s_\theta c_\psi - c_\phi s_\psi & c_\theta c_\psi \end{bmatrix}^\top
\begin{bmatrix} \sum_{i=1}^{4} F_i \\ 0 \end{bmatrix} -
\begin{bmatrix} 0 \\ 0 \\ g \end{bmatrix}, \\
\begin{bmatrix} \dot{\psi} \\ \dot{\theta} \\ \dot{\phi} \end{bmatrix} &=
- \begin{bmatrix} a_\psi & 0 & 0 \\ 0 & a_\theta & 0 \\ 0 & 0 & a_\phi \end{bmatrix}
\begin{bmatrix} \dot{\psi} \\ \dot{\theta} \\ \dot{\phi} \end{bmatrix} +
\begin{bmatrix} \frac{\ell}{J_\psi} & 0 & 0 \\ 0 & \frac{\ell}{J_\theta} & 0 \\ 0 & 0 & \frac{C}{J_\phi} \end{bmatrix}
\begin{bmatrix} -1 & 1 & 1 & -1 \\ -1 & -1 & 1 & 1 \\ 1 & -1 & 1 & -1 \end{bmatrix}
\begin{bmatrix} F_1 \\ F_2 \\ F_3 \\ F_4 \end{bmatrix},
\end{align*}

where \(x, y, z\) are the Cartesian coordinates of the centre of mass; \(\phi, \theta, \psi\) are the roll, pitch, and yaw angles; \(g = 9.81 \, \text{m/s}^2\) is the gravitational acceleration; \(m\) is the mass; \(F_i,\ i = 1,\dots,4\) are the rotor thrust forces; \(\ell\) is the distance between the centre of mass and the rotors; \(J_\psi, J_\theta, J_\phi\) are the moments of inertia; \(C\) is the proportionality coefficient; \(a_x, a_y, a_z, a_\psi, a_\theta, a_\phi\) are the viscous friction coefficients; \(c(\cdot) \equiv \cos(\cdot)\); \(s(\cdot) \equiv \sin(\cdot)\).

Choose the signal for the rotor thrust forces \(F_i\) as

\[
\begin{bmatrix} F_1 \\ F_2 \\ F_3 \\ F_4 \end{bmatrix} =
\begin{bmatrix} 1 & 1 & -1 & -1 \\ 1 & -1 & 1 & -1 \\ 1 & 1 & 1 & 1 \\ 1 & -1 & -1 & 1 \end{bmatrix}
\begin{bmatrix} m & 0 & 0 & 0 \\ 0 & \frac{J_\phi}{C} & 0 & 0 \\ 0 & 0 & \frac{J_\psi}{\ell} & 0 \\ 0 & 0 & 0 & \frac{J_\theta}{\ell} \end{bmatrix}
\begin{bmatrix} u_1 + g \\ u_2 \\ u_3 \\ u_4 \end{bmatrix}
\]

and with the change of variables

\begin{align*}
\tilde{\xi}_1 &=
\begin{bmatrix} \tilde{\xi}_{11} \\ \tilde{\xi}_{12} \end{bmatrix}
=
\begin{bmatrix} z - z^* \\ \phi - \phi^* \end{bmatrix}
=
\xi_1 -
\begin{bmatrix} z^* \\ \phi^* \end{bmatrix}, &
\tilde{\xi}_2 &= \xi_2, \\
\tilde{\xi}_3 &=
\begin{bmatrix} \tilde{\xi}_{31} \\ \tilde{\xi}_{32} \end{bmatrix}
=
\begin{bmatrix} x - x^* \\ y - y^* \end{bmatrix}
=
\xi_3 -
\begin{bmatrix} x^* \\ y^* \end{bmatrix}, &
\tilde{\xi}_4 &= \xi_4, \\
\tilde{\xi}_5 &= \xi_5, &
\tilde{\xi}_6 &= \xi_6,
\end{align*}

and using Proposition 1 from [3], we obtain the dynamic model of the quadcopter motion in deviations from the desired position and orientation:

\begin{align}
\dot{\tilde{\xi}}_1 &= \tilde{\xi}_2, \notag \\
\dot{\tilde{\xi}}_2 &= q_1(\psi, \theta) + b_1(\psi, \theta) \begin{bmatrix} u_1 \\ u_2 \end{bmatrix}, \notag \\
\dot{\tilde{\xi}}_3 &= \tilde{\xi}_4, \notag \\
\dot{\tilde{\xi}}_4 &= \beta(t)\tilde{\xi}_5, \notag \\
\dot{\tilde{\xi}}_5 &= \tilde{\xi}_6, \\
\dot{\tilde{\xi}}_6 &= q_2(\phi, \psi, \theta, \dot{\phi}, \dot{\psi}, \dot{\theta}) + b_{21}(\phi, \psi, \theta)u_2 + b_{22}(\phi, \psi, \theta) \begin{bmatrix} u_3 \\ u_4 \end{bmatrix}. \notag
\end{align}

The control objective is to synthesize an output feedback control law for \(u_1, u_2, u_3, u_4\) such that the equilibrium \(\tilde{\xi} = \col(\tilde{\xi}_1, \tilde{\xi}_2, \tilde{\xi}_3, \tilde{\xi}_4, \tilde{\xi}_5, \tilde{\xi}_6) = 0\) is asymptotically stable.

\section{State Feedback Control Synthesis}

\textbf{Step 1.} Introduce new aggregated variables

\[
\xi_1 = \begin{bmatrix} \tilde{\xi}_1 \\ \tilde{\xi}_3 \end{bmatrix}, \quad \xi_2 = \begin{bmatrix} \tilde{\xi}_2 \\ \tilde{\xi}_4 \end{bmatrix},
\]

then the complete quadcopter motion model (1)--(2) takes the form

\begin{align}
\dot{\xi}_1 &= \xi_2, \notag \\
\dot{\xi}_2 &= \begin{bmatrix} q_1(\psi, \theta) \\ 0 \end{bmatrix} + \begin{bmatrix} b_1(\psi, \theta) & 0 \\ 0 & \beta(t)I_2 \end{bmatrix} \begin{bmatrix} u_1 \\ u_2 \\ \xi_5 \end{bmatrix}, \notag \\
\dot{\xi}_5 &= \tilde{\xi}_6, \\
\dot{\xi}_6 &= q_2(\phi, \psi, \theta, \dot{\phi}, \dot{\psi}, \dot{\theta}) + b_{21}(\phi, \psi, \theta)u_2 + b_{22}(\phi, \psi, \theta) \begin{bmatrix} u_3 \\ u_4 \end{bmatrix}. \notag
\end{align}

\textbf{Step 2.} Let the control vector \(\begin{bmatrix} u_1 \\ u_2 \end{bmatrix}\) be the output of two integrators with an input \(\begin{bmatrix} v_1 \\ v_2 \end{bmatrix}\) to be designed later:

\[
\dot{u}_{12} = \rho_{12}, \quad \dot{\rho}_{12} = v_{12}.
\]

\textbf{Step 3.} The aggregated quadcopter motion model can be represented in a normal form.

\noindent \textbf{Proposition 1.} The aggregated quadcopter motion model can be represented in the normal form

\begin{align}
\dot{\zeta}_1 &= \zeta_2, \notag \\
\dot{\zeta}_2 &= \zeta_3, \notag \\
\dot{\zeta}_3 &= \zeta_4, \\
\dot{\zeta}_4 &= q_4(\zeta, \phi) + b_4(\zeta, \phi)U, \notag
\end{align}

where \(\zeta = \col(\zeta_1, \zeta_2, \zeta_3, \zeta_4) \in \mathbb{R}^{16}\) is the state vector, \(U = \col(v_1, v_2, u_3, u_4)\) is the control input vector, and the functions \(q_4(\zeta, \phi)\) and \(b_4(\zeta, \phi)\) satisfy the following properties:

\[
q_4(0, \phi) = 0,
\]

the matrix \(b_4(\zeta, \phi)\) is nonsingular for all values of its arguments, and

\[
b_4(0, \phi) = g \begin{bmatrix} 0 & 0 \\ 0 & 0 \\ 1 & 0 \\ 0 & 1 \end{bmatrix}, \quad \beta \begin{bmatrix} -c_\phi & -s_\phi \\ -s_\phi & c_\phi \end{bmatrix}.
\]

\noindent \textit{Proof} of Proposition~1 can be obtained by successively computing the derivatives of the variable \(\zeta_3 = \dot{\zeta}_2\) and \(\zeta_4 = \dot{\zeta}_3\).

For system (4), the state feedback control law takes the form

\begin{equation}
U = \begin{bmatrix} v_1 \\ v_2 \\ u_3 \\ u_4 \end{bmatrix} = b_4(\zeta, \phi)^{-1}\bigl[-q_4(\zeta, \phi) - \gamma_1\zeta_1 - \gamma_2\zeta_2 - \gamma_3\zeta_3 - \gamma_4\zeta_4\bigr],
\end{equation}

where the choice of gains \(\gamma_1, \gamma_2, \gamma_3, \gamma_4 > 0\) is determined by the desired performance of the closed-loop system, whose model becomes

\[
\begin{bmatrix} \dot{\zeta}_1 \\ \dot{\zeta}_2 \\ \dot{\zeta}_3 \\ \dot{\zeta}_4 \end{bmatrix} = \begin{bmatrix} 0 & I_4 & 0 & 0 \\ 0 & 0 & I_4 & 0 \\ 0 & 0 & 0 & I_4 \\ -\gamma_1I_4 & -\gamma_2I_4 & -\gamma_3I_4 & -\gamma_4I_4 \end{bmatrix}
\begin{bmatrix} \zeta_1 \\ \zeta_2 \\ \zeta_3 \\ \zeta_4 \end{bmatrix},
\]

from which it is easy to see that the parameters \(\gamma_1, \gamma_2, \gamma_3, \gamma_4\) can be chosen as the corresponding coefficients of standard Butterworth or Newton characteristic polynomials, or computed via modal control methods.

In accordance with Step~2 of the above algorithm, for system (3) the control law (6) is supplemented by

\[
\begin{bmatrix} \dot{u}_1 \\ \dot{u}_2 \end{bmatrix} = \begin{bmatrix} v_1 \\ v_2 \end{bmatrix},
\]

however, the practical implementation of the state feedback law (6) is hampered because it requires knowledge of the full state \(\zeta = \col(\zeta_1, \zeta_2, \zeta_3, \zeta_4)\) and of the functions \(b_4(\zeta, \phi)^{-1}\) and \(q_4(\zeta, \phi)\). Therefore, the next section addresses the synthesis of an output feedback controller.

\section{Output Feedback Control Synthesis}

Following [4], for system (4) we choose the output feedback control law

\begin{equation}
U = \begin{bmatrix} v_1 \\ v_2 \\ u_3 \\ u_4 \end{bmatrix} = \sat_N\Bigl(b_4(0, \phi)^{-1}\bigl[-\sigma - \gamma_1\hat{\zeta}_1 - \gamma_2\hat{\zeta}_2 - \gamma_3\hat{\zeta}_3 - \gamma_4\hat{\zeta}_4\bigr]\Bigr),
\end{equation}

where \(\sat_N(\cdot)\) is a smooth saturation function with level \(N\), \(b_4(0, \phi)\) is the matrix from (5), and \(\sigma\), \(\hat{\zeta}\) are the states of an extended observer of the form

\begin{align}
\dot{\hat{\zeta}}_1 &= \hat{\zeta}_2 + \kappa A_4(\zeta_1 - \hat{\zeta}_1), \notag \\
\dot{\hat{\zeta}}_2 &= \hat{\zeta}_3 + \kappa^2 A_3(\zeta_1 - \hat{\zeta}_1), \notag \\
\dot{\hat{\zeta}}_3 &= \hat{\zeta}_4 + \kappa^3 A_2(\zeta_1 - \hat{\zeta}_1), \\
\dot{\hat{\zeta}}_4 &= \sigma + b_4(0, \phi)U + \kappa^4 A_1(\zeta_1 - \hat{\zeta}_1), \notag \\
\dot{\sigma} &= \kappa^5 A_0(\zeta_1 - \hat{\zeta}_1), \notag
\end{align}

where \(\kappa\) is a high gain, and \(A_0, A_1, A_2, A_3, A_4\) are positive definite matrices such that the eigenvalues of the matrix

\[
A = \begin{bmatrix} -A_4 & I_4 & 0 & 0 & 0 \\ -A_3 & 0 & I_4 & 0 & 0 \\ -A_2 & 0 & 0 & I_4 & 0 \\ -A_1 & 0 & 0 & 0 & I_4 \\ -A_0 & 0 & 0 & 0 & 0 \end{bmatrix}
\]

are real and negative.

It can be shown that the control law (7), (8) guarantees semi-global asymptotic stability of the closed-loop system. Importantly, the implementation of this controller does not require knowledge of the full state \(\zeta = \col(\zeta_1, \zeta_2, \zeta_3, \zeta_4)\) nor of the functions \(b_4(\zeta, \phi)^{-1}\) and \(q_4(\zeta, \phi)\).

\section{Conclusion}

In this work, the problem of output feedback control for quadcopter motion has been solved. The solution steps include model transformations to bring the system into a normal form, synthesis of a state feedback control law that achieves feedback linearization, and finally, synthesis of an output feedback control law that ensures semi-global asymptotic stability of the closed-loop system.

\vspace{0.3cm}
\noindent \textit{This work was supported by the Ministry of Science and Higher Education of the Russian Federation (state assignment passport No.~2019-089).}


\begin{thebibliography}{99}
\bibitem{1} Borisov O. I., Pyrkin A. A., Isidori A. Application of enhanced extended observer in stationkeeping of a quadrotor with unmeasurable pitch and roll angles // IFAC-PapersOnLine. 2019. Vol.~52, No.~16. P.~837--842.
\bibitem{2} Borisov O. I., Kakanov M. A., Zhivitskii A. Yu., Pyrkin A. A. Robust output trajectory control of a quadcopter based on a geometric approach // Journal of Instrument Engineering. 2021. Vol.~64, No.~12. P.~982--992. (In Russian)
\bibitem{3} Kim S. A., Pyrkin A. A., Borisov O. I. Control algorithms for quadcopter motion in dynamic positioning mode // Journal of Instrument Engineering. 2023. Vol.~66, No.~10. P.~834--844. (In Russian)
\bibitem{4} Freidovich L. B., Khalil H. K. Performance recovery of feedback-linearization-based designs // IEEE Transactions on Automatic Control. 2008. Vol.~AC-53, No.~10. P.~2324--2334.
\end{thebibliography}
\end{document}